\documentclass[aps, prd, reprint, nofootinbib, longbibliography, preprintnumbers, amsfonts, amsmath, amssymb]{revtex4-1}

\usepackage{ifpdf}
\usepackage{physics}
\usepackage[utf8]{inputenc}

\ifpdf
    \usepackage[pdftex, colorlinks = true,plainpages = false, pdfpagelabels]{hyperref}
    \usepackage[pdftex]{color}
    \hypersetup{colorlinks = true}
\else
    \usepackage[colorlinks = true]{hyperref}
\fi

\begin{document}

\title{Oscillating and Static Universes from a Single Barotropic Fluid}

\author{John Kehayias}
\email{john.kehayias@vanderbilt.edu}
\affiliation{Department of Physics and Astronomy, Vanderbilt University\\
  Nashville, TN 37235, United States}
\author{Robert J. Scherrer}
\email{robert.scherrer@vanderbilt.edu}
\affiliation{Department of Physics and Astronomy, Vanderbilt University\\
  Nashville, TN 37235, United States}

\begin{abstract}

  \noindent
  We consider cosmological solutions to general relativity with a
  single barotropic fluid, where the pressure is a general function of
  the density, $p = f(\rho)$. We derive conditions for static and
  oscillating solutions and provide examples, extending earlier work
  to these simpler and more general single-fluid cosmologies.
  Generically we expect such solutions to suffer from instabilities,
  through effects such as quantum fluctuations or tunneling to zero
  size. We also find a classical instability (``no-go'' theorem) for
  oscillating solutions of a single barotropic perfect fluid due to a
  necessarily negative squared sound speed.

\end{abstract}

\date{\today}

\maketitle

\section{Introduction}
\label{sec:intro}

The broad dynamics of the universe can be categorized generally as
expanding, contracting, or static. The current state, as well as we
understand it, is that of accelerating expansion. However, this does
not preclude such dynamics being a part of an overall cyclical
behavior, or having an origin in a static phase. Such cosmological
solutions to General Relativity (GR) have been known and studied for
many years, and are still a current research interest.

Besides the novelty of eternal static and oscillating universes, there
are several cosmological (and perhaps metaphysical) questions they
attempt to answer. For any (past) eternal universe, the question of
the origin of time is no longer relevant as there is no ``beginning''
(see, e.g.~\cite{Mithani:2012ii}). On the other hand, one is instead
faced with the question of how a static universe becomes the dynamical
universe we observe. Oscillating universes are a framework to answer
both the history and future of the universe, and often aim to repeat
forever. Both of these types of universes seek to evade an initial
singularity, such as the Big Bang.

However, general results of singularity
theorems~\cite{singularity_theorem1, *singularity_theorem2,
  *singularity_theorem3, *singularity_theorem4, *singularity_theorem5,
  *singularity_theorem6} in GR prove that most cosmologies that have
been studied unavoidably have some sort of singularity (i.e.~geodesics
necessarily have a starting point). As with many no-go type theorems,
it is the assumptions and loopholes that spur further creativity on
the part of theorists and model builders.

One such exception is the Einstein static universe, which is the
starting point for the ``emergent universe'' inflationary
scenario~\cite{emergent_universe1,*emergent_universe2}. This solution
evades the singularity theorems by being a closed (curvature constant
$k = +1$), static (Hubble parameter $H = 0$) universe. While this
purports to be a (classical) solution with no beginning of time, it is
classically unstable to homogeneous linear perturbations (necessary
for a transition to inflation) and at best neutrally stable against
inhomogeneous perturbations~\cite{es_stability0, *es_stability1,
  *es_stability2, *es_stability3, *es_stability4}, as well as
suffering from other quantum instabilities. Two such problems are in
the precise tuning needed for the static solution, which will be
broken quantum mechanically~\cite{Aguirre:2013kea}, as well as the
possibility of tunneling to zero size (``tunneling to
nothing''~\cite{Mithani:2011en}). As such scenarios are enticing to
many because of their eternal nature, even the tiniest probability of
a such behavior is disastrous.

Oscillating universes (see, e.g.~\cite{Dabrowski:1995ae}) are another
type of interesting model that can avoid the singularity theorems. A
study of such solutions has been done systematically in the
past~\cite{Harrison01091967, *Novello:2008ra}, given various
assumptions and ingredients. More recently, Graham et
al.~\cite{Graham:2011nb, *Graham:2014pca} have explored a new class of
oscillating models, which they have dubbed the ``simple harmonic
universe.'' In these models, a classically stable oscillating solution
is achieved through a combination of positive curvature, a negative
cosmological term, and a fluid with an equation of state parameter
satisfying $-1 < w < -1/3$, where the equation of state parameter $w$
for a fluid with density $\rho$ and pressure $p$ is defined as
\begin{equation}
\label{eq:w}
w = p/\rho
\end{equation}
Note, however, that this class of models is also unstable to tunneling
to zero size~\cite{Mithani:2011en} (there may also be an instability
due to quantum particle production, as in oscillations of an Einstein
static solution~\cite{Bag:2014tta}).

With such a wide range of inputs, such as exotic matter, not all of
the possibilities have been exhausted, even in standard GR\@. In this
work we explore static and oscillating solutions (with positive
curvature) for a very general equation of state, the so-called
``barotropic fluid,'' which encapsulates many different models. We
analyze both the general conditions for these types of solutions, as
well as toy examples, simplifying and generalizing previous studies.
In the next section, we examine barotropic models and derive the
conditions necessary to achieve either a stable static solution or an
oscillating universe. While we find (classically) stable static
solutions, there is a classical instability for oscillating solutions
with a single barotropic perfect fluid. We also show how the simple
harmonic universe can be mapped onto our model. In
Sec.~\ref{sec:instab}, we examine quantum instabilities in our model.
We show that these instabilities generically also apply to the
barotropic models, further extending the analysis that one cannot
easily construct such infinitely long-lived cosmologies. Our
conclusions are summarized in Sec.~\ref{sec:discon}.

\section{Barotropic Cosmology}
\label{sec:baro}

In this section we will introduce the basics of barotropic fluids and
their solutions in GR\@. By a ``barotropic fluid'' we mean some matter
(or fields) with pressure $p$ and density $\rho$, with the equation of
state given by
\begin{equation}
  \label{eq:barodef}
  p = f(\rho).
\end{equation}
Barotropic fluids have long been studied in the context of dark
energy. Specific examples include the well-known Chaplygin gas
\cite{Kamenshchik, *Bilic} and the generalized Chaplygin gas
\cite{Bento}, the linear (affine) equation of state~\cite{linear1,
  *linear2, *Quercellini, AB}, the quadratic equation of state
\cite{Nojiri:2004pf, AB}, and the Van der Waals equation of
state~\cite{VDW1, *VDW2} (see also the review~\cite{Bamba:2012cp}). A
general study of the properties of barotropic models for dark energy
was undertaken in Refs.~\cite{LinderScherrer, Bielefeld}. It is often
convenient to define an equation of state parameter $w$, given by Eq.
(\ref{eq:w}). So, for instance, nonrelativistic matter is
characterized by $w=0$, while a pure cosmological constant has
$w = -1$.

Assuming an isotropic and homogeneous universe, we have the Friedmann
equations,
\begin{align}
  \label{eq:friedman1}
  \frac{\ddot{a}}{a} = -\frac{1}{6}\left( \rho + 3p \right),\\
  \label{eq:friedman2}
  \left(\frac{\dot a}{a}\right)^2 = -\frac{k}{a^2} + \frac{1}{3}\rho,
\end{align}
where $a$ is the scale factor, $k = \pm1, 0$ is the curvature
parameter, and we work in units where
$\hbar = c = k_\text{B} = 8\pi G = 1$. For a barotropic fluid with
equation of state parameter $w$, we can rewrite Eq.
(\ref{eq:friedman1}) as
\begin{equation}
  \label{eq:friedman1w}
  \frac{\ddot{a}}{a} = -\frac{1}{6}\rho \left(1 + 3w \right)
\end{equation}
We also have the equation for the evolution of the energy density,
\begin{equation}
  \label{eq:rhoevol}
  \dv{\ln \rho}{\ln a} = -3(1+w).
\end{equation}

\subsection{Static Solutions}
\label{sec:static}
Consider first the case of static, stable solutions to the Friedman
equations for a universe containing a single barotropic fluid with
equation of state given by Eq. (\ref{eq:barodef}). A static solution
requires $\dot a = \ddot a = 0$ at some fixed density $\rho = \rho_*$
and scale factor $a = a_*$. Then Eq.~\eqref{eq:friedman1w} immediately
tells us that $\ddot a = 0$ requires $w = -1/3$. To achieve
$\dot a=0$, we need (from Eq.~\eqref{eq:friedman2})
\begin{equation}
  \label{eq:rhostatic}
  \rho_* = \frac{3k}{a_*^2},
\end{equation}
which requires a universe with positive curvature
($k = +1$).

While these conditions are sufficient for a static solution, they do
not insure stability. For a stable solution, we require that a small
increase in $a$ away from the stable solution yields $\ddot a < 0$,
while a small decrease gives $\ddot a > 0$. From
Eq.~\eqref{eq:friedman1w}, this stability condition will be satisfied
as long as
\begin{equation}
  \label{eq:dwda}
  \dv{w}{a} > 0.
\end{equation}

How does this stability condition translate into a constraint on the
equation of state $f(\rho)$? For a barotropic fluid, it is easy to
verify that~\cite{LinderScherrer}
\begin{equation}
  \label{eq:LS}
  a \dv{w}{a} = -3(1+w)\left(\dv{p}{\rho} - w \right)
\end{equation}
Stability requires the right-hand side of this expression to be
positive, so, with $w(a_*) = -1/3$,
\begin{equation}
  \dv{p}{\rho} < -1/3.
\end{equation}

As an example, consider the generalized Chaplygin gas. The equation of
state is given by~\cite{Bento}
\begin{equation}
  \label{eq:Chaplygin}
  p = -\frac{A}{\rho^\alpha},
\end{equation}
corresponding to a density evolution
\begin{equation}
  \label{eq:rhoChap}
  \rho/\rho_0 = \left[A_s + (1-A_s) (a/a_0)^{-3(1+\alpha)} \right]^{1/(1+\alpha)},
\end{equation}
where the $0$ subscript denotes quantities evaluated at an arbitrary
fiducial value of the scale factor, and $A_s$ is given by
\begin{equation}
  A_s = A/\rho_0^{1+\alpha}.
\end{equation}
The equation of state parameter is given by
\begin{equation}
  \label{eq:wChap}
  w = - \frac{A_s}{A_s + (1-A_s)(a/a_0)^{-3(1+\alpha)}}.
\end{equation}

The original version of this model assumed that $\alpha > -1$ and
$A_s < 1$, causing the Chaplygin gas to behave like nonrelativistic
matter at early times and a cosmological constant at late times.
However, Sen and Scherrer~\cite{SenScherrer} extended this model to
$\alpha < -1$ and $A_s >1$. Setting $w=-1/3$ in Eq.~\eqref{eq:wChap}
we obtain the condition
\begin{equation}
  \label{eq:AsChap}
  A_s = \frac{1}{1+2 (a_*/a_0)^{3(1+\alpha)}},
\end{equation}
which implies that $0 < A_s < 1$. Then from Eq.~\eqref{eq:wChap}, we
see that the stability condition, $\dv*{w}{a} > 0$, gives
$\alpha < -1$. These parameters for the generalized Chaplygin gas
allow for a stable, static solution.

\subsection{Oscillating Solutions}
\label{sec:oscillate}
The existence of a stable, static solution automatically implies the
existence of oscillating solutions, since one can simply perturb
around the stable scale factor. However, we can derive a more general
set of oscillating solutions that do not require small perturbations.

Assume that a solution exists for which the universe is oscillating
between $a_\text{min}$ and $a_\text{max}$. Then $\dot a = 0$ at both
$a_\text{min}$ and $a_\text{max}$, which will be satisfied as long as
the density at these two scale factors is given by
\begin{equation}
  \rho_{\substack{\text{min,}\\\text{max}}} = \frac{3k}{a_{\substack{\text{min,}\\\text{max}}}^2}.
\end{equation}

The oscillating solution also requires $\ddot a < 0$ at
$a = a_\text{max}$ and $\ddot a > 0$ at $a = a_\text{min}$. This will
be achieved when the equation of state parameter for the barotropic
fluid obeys
\begin{align}
  w(a_\text{max}) &> -1/3,\\
  w(a_\text{min}) &< -1/3.
\end{align}
If $w$ is a monotonic function of $a$, then these two equations imply
that
\begin{equation}
  \dv{w}{a} > 0,
\end{equation}
during both the expanding and contracting phases, just as in the case
of the static solution.

Now we can translate this into a constraint on the pressure as a function of the density. From
equation \eqref{eq:LS}, the requirement that $\dv*{w}{a} > 0$ is
equivalent to the condition
\begin{equation}
  \label{eq:dpdrhobound}
  \dv{p}{\rho} < \frac{p}{\rho}.
\end{equation}
For the special case where $w$ always remains negative, this condition
reduces to the particularly simple form:
\begin{equation}
  \label{eq:lncondition}
  \dv{\ln p}{\ln \rho} > 1.
\end{equation}

We can again take the generalized Chaplygin gas as a specific example.
For the Chaplygin gas equation of state given by
Eq.~\eqref{eq:Chaplygin}, our condition in Eq.~\eqref{eq:lncondition}
will be satisfied as long at $\alpha < -1$. The general behavior of
such models was examined in Ref.~\cite{SenScherrer}. At small $a$, the
equation of state parameter approaches $w = -1$, while at large $a$ it
asymptotically behaves like pressureless dust ($w \rightarrow 0$).
Hence, these models were dubbed ``transient generalized Chaplygin
gas'' models. However, in the presence of a positive curvature, this
type of fluid can generate oscillating solutions.

As was the case for the oscillating solutions discussed in
Refs.~\cite{Graham:2011nb, *Graham:2014pca}, these models suffer from
a classical instability. If we assume that our barotropic fluid is a
perfect fluid, than perturbation growth will be unstable whenever the
sound speed $c_s^2 \equiv \dv*{p}{\rho} < 0$ (see, e.g.,
\cite{LinderScherrer}). However, the upper bound on $c_s^2$ given by
Eq.~\eqref{eq:dpdrhobound}, combined with the requirement the
$w < -1/3$ at $a_\text{min}$ automatically produces a negative
$c_s^2$. This can be taken as a ``no-go'' theorem: {\it no} barotropic
perfect fluid can combine with positive curvature to produce an
oscillating universe. However, as noted in Refs.~\cite{Graham:2011nb,
  *Graham:2014pca}, it is possible to find non-perfect fluid models
that mimic a particular equation of state, but with a different sound
speed.

\subsection{Comparison with the Simple Harmonic Universe}
Here we show that the background (homogeneous) evolution of the simple
harmonic universe proposed in Refs.~\cite{Graham:2011nb,
  *Graham:2014pca} can be put into the context of a single barotropic
fluid, giving a simpler harmonic universe. This section will also
serve as as a guideline to show how multi-component models can be put
into the context of a model with a single barotropic fluid.

The ingredients proposed in the simple harmonic universe are positive
curvature, a positive cosmological constant, and fluid with
$w = -2/3$. From Eq.~\eqref{eq:rhoevol}, the density of the fluid
component scales as $\rho \propto a^{-1}$, so the total density and
total pressure in this model are
\begin{equation}
  \rho = \rho_0\left(\frac{a_0}{a}\right) + \rho_\Lambda,
\end{equation}
and
\begin{equation}
  p = -\frac{2}{3} \rho_0 \left(\frac{a_0}{a}\right) - \rho_\Lambda,
\end{equation}
where $\rho_\Lambda$ and $\rho_0$ are constant. These expressions for
pressure and density can then be combined to yield the corresponding
single-fluid equation of state
\begin{equation}
  p = -\frac{2}{3} \rho - \frac{1}{3} \rho_\Lambda.
\end{equation}
This is a form of the linear/affine equation of state previously
studied in Refs.~\cite{linear1, *linear2, *Quercellini, AB}. Note
that it satisfies (as it must) Eq.~\eqref{eq:dpdrhobound} as long as
$\rho_\Lambda > 0$.

\section{Quantum Instabilities}
\label{sec:instab}
While such static (and perhaps oscillating) solutions may be
classically stable (or at least not unstable), the situation is more
complicated once we include quantum mechanical effects.

One possibility is that the universe can tunnel to another state.
Specifically, we will be concerned with tunneling to zero size,
$a \rightarrow 0$, or ``tunneling to nothing.'' To calculate this
process we will use the Wheeler-DeWitt equation (we will
follow~\cite{Mithani:2011en}; see also~\cite{quantumcosmoreview} for a
review). We start with the classical Hamiltonian for the Friedmann
equation, Eq.~\eqref{eq:friedman2},
\begin{equation}
  \label{eq:hclass}
  H = -\frac{1}{24\pi^2 a}\left(p_a^2 + U(a)\right),
\end{equation}
with the momentum conjugate,
\begin{equation}
  \label{eq:pa}
  p_a \equiv -12\pi^2 a\dot{a},
\end{equation}
and potential,
\begin{equation}
  \label{eq:wdwpot}
  U(a) = (12\pi^2 a)^2\left(k - \frac{1}{3}a^2\rho(a)\right).
\end{equation}
With the canonical quantization of $p_a \rightarrow -i\dv*{a}$, the Hamiltonian
becomes the operator $\mathcal{H}$ acting on $\psi$, the so-called
wavefunction of the universe,
\begin{equation}
  \label{eq:wdw}
  \mathcal{H}\psi = 0.
\end{equation}

We will use the generalized Chaplygin gas as our prototypical example,
namely the density evolution of Eq.~\eqref{eq:rhoChap} (with $k = 1$).
The potential then has the form
\begin{align}
  \label{eq:potChap}
  U(a) = &(12\pi^2a)^2\left(\rule{0cm}{0.77cm}1 - \frac{1}{3}a^2\rho_0 \Biggl[A_s \right.\Biggr.\nonumber\\
  &\left.\left.+ (1 - A_s)\left(\frac{a}{a_ 0}\right)^{-3(1 + \alpha)}\right]^{1/(1+\alpha)} \right).
\end{align}

Immediately we see that, with $\alpha < -1$, generically there is a minimum
at $a = 0$.\footnote{To avoid a minimum at $a = 0$ the density would
  need a \textit{negative} component that goes to zero faster than
  $a^{-2}$.} A second minimum at finite, nonzero $a$ must be computed
given the parameters $A_s$ and $\alpha$. We can compute a probability to
collapse to $a = 0$ using the WKB action,
\begin{equation}
  \label{eq:wkbs}
  S_\text{WKB} = \int_0^{a_-}\dd{a} \sqrt{U(a)},
\end{equation}
with $U(a_-) = 0$ the turning point closest to $a = 0$. The probability
is proportional to $\exp(-2S_\text{WKB})$. These calculations can be
performed numerically, but it is clear that even a tiny probability
for tunneling to zero size is disastrous if one wants an eternal
universe.

One may wonder how general such a problem might be and possible ways
to escape such a fate (e.g.~using the Casimir
energy~\cite{Graham:2014pca, casimir}). One approach would be to
impose a boundary condition at $a = 0$ such that the wavefunction
vanishes. However, there is no such freedom once the boundary
condition that the wavefunction $\psi \rightarrow 0$ as
$a \rightarrow \infty$ is required~\cite{casimir}. Thus, in general
one cannot evade such a tunneling process, but in special
circumstances the wavefunction may nonetheless be zero at
$a = 0$~\cite{casimir}.

However, such an application for the Wheeler-DeWitt equation is
perhaps overreaching in its applicability. We would expect that
quantum gravity effects (from a complete theory of quantum gravity) to
become important as the size of the universe approaches the Planck
length, let alone smaller. So while tunneling to nothing appears to be
legitimate concern for these types of models, we cannot say
definitively that such a calculation is valid.

There are also other quantum instabilities that may arise in these
models due to quantum fluctuations. Consider the static solution,
which requires a specific value for the energy density, given the
scale factor, Eq.\eqref{eq:rhostatic}. Quantum fluctuations may upset
such a balance (e.g.~for an analysis for such a problem with the
emergent universe scenario, see~\cite{Aguirre:2013kea}). Furthermore,
the universe cannot truly be static forever, as eventually there needs
to be some sort of evolution, such as inflation. There is then a
general conflict between something that is stable indefinitely and yet
eventually ``does something.'' Thus eternal universes with a static
``beginning'' are contradictory by their nature.

\section{Discussion and Conclusions}
\label{sec:discon}

Our results show that both static and oscillating behavior can be
achieved in the context of a universe with positive curvature
containing a single barotropic fluid, and we have derived the
corresponding conditions on such a fluid. Note that there is a simple
mapping from barotropic fluids to purely kinetic $k$-essence
models~\cite{LinderScherrer}, so our results are easily generalized to
the latter class of models. Unfortunately, the quantum instabilities
that plague previous models of this kind also apply to our model as
well.

In this work we have focused on models which classically have no
singularities. It is possible to avoid the barotropic ``no-go''
theorem by loosening this restriction, e.g.~a singularity in $H$ as in
the ``sneezing universe''~\cite{Tilquin:2015gza}. Of course, in this
case one loses one of the most appealing features of static and
oscillating models (lack of singularities) while also introducing
further difficulties and assumptions in the model.

The static and oscillating solutions with a single barotropic fluid
are a generalization and extension of previous studies of these types
of solutions in GR\@. Our exploration of barotropic oscillating
cosmologies found a new difficulty in constructing such models, in
addition to previous quantum instabilities. Thus we have also found
that it is difficult to find truly eternal models. How this may fit in
to answering broad questions, such as if the universe necessarily had
a beginning, is an open question.

\begin{acknowledgments}
  \noindent
  J.K.~and R.J.S.~were supported in part by the DOE (DE-SC0011981).
\end{acknowledgments}

\clearpage

\bibliography{barotropic_refs}

\end{document}